\documentclass[sigconf]{acmart}

\AtBeginDocument{%
  }

\setcopyright{acmlicensed}

\copyrightyear{2025}
\acmYear{2025}
\acmDOI{10.1145/XXXXXXX.XXXXXXX}

\acmConference[UIST '25 Poster]{Adjunct Proceedings of the 38th Annual ACM Symposium on User Interface Software and Technology – Poster}{September 28--October 1, 2025}{Busan, Republic of Korea}
\acmISBN{978-1-4503-XXXX-X/25/09}

\begin{document}

%%
%% The "title" command has an optional parameter,
%% allowing the author to define a "short title" to be used in page headers.
\title{\emph{MeloKids}: Multisensory VR System to Enhance Speech and Motor Coordination in Children with Hearing Loss}

%%
%% The "author" command and its associated commands are used to define
%% the authors and their affiliations.
%% Of note is the shared affiliation of the first two authors, and the
%% "authornote" and "authornotemark" commands
%% used to denote shared contribution to the research.
\author{Yichen Yu}
\authornote{Both authors contributed equally to this research.}
\email{lunarsboy@gmail.com}
\orcid{0009-0001-0175-3253}

\affiliation{%
  \institution{AYXR}
  \institution{North Carolina State University}
  \city{Raleigh}
  \state{North Carolina}
  \country{USA}
}
\affiliation{%
  \institution{Georgia Institute of Technology}
  \city{Atlanta}
  \state{Georgia}
  \country{USA}
}

\author{Qiaoran Wang}
\authornotemark[1]
\email{violetforever999@gmail.com}

\affiliation{%
  \institution{AYXR}
  \institution{National University of Singapore}
  \country{Singapore}
}

%%
%% By default, the full list of authors will be used in the page
%% headers. Often, this list is too long, and will overlap
%% other information printed in the page headers. This command allows
%% the author to define a more concise list
%% of authors' names for this purpose.
\renewcommand{\shortauthors}{Yu and Wang}

%%
%% The abstract is a short summary of the work to be presented in the
%% article.
\begin{abstract}
Children with hearing impairments face ongoing challenges in language and motor development. This study explores how multi-sensory feedback technology based on virtual reality (VR), integrating auditory, visual, and tactile stimuli, can enhance rehabilitation outcomes. Using functional near-infrared spectroscopy (fNIRS) technology, we assessed cortical activation patterns in children during pitch-matching tasks across different interaction modes. Our findings aim to provide evidence for designing personalized, interactive rehabilitation systems that enhance cognitive engagement and motor control in children with hearing impairments.
\end{abstract}

%%
%% The code below is generated by the tool at http://dl.acm.org/ccs.cfm.
%% Please copy and paste the code instead of the example below.
%%



%%
%% Keywords. The author(s) should pick words that accurately describe
%% the work being presented. Separate the keywords with commas.
%% A "teaser" image appears between the author and affiliation
%% information and the body of the document, and typically spans the
%% page.

%%
%% This command processes the author and affiliation and title
%% information and builds the first part of the formatted document.
\maketitle

\section{INTRODUCTION}
Children with hearing impairments often experience significant delays in language development and motor coordination, even after receiving cochlear implants (CIs). While cochlear implants significantly improve speech perception in many individuals, their ability to distinguish pitch and temporal cues remains limited due to restricted stimulus channels. As a result, children with hearing impairments continue to face challenges in speech modulation, pitch accuracy, and rhythmic synchrony—core elements of expressive language and communication. These difficulties are often exacerbated by weakened auditory-vocal feedback loops, which play a critical role in speech self-monitoring and learning \cite{amir2003effect, lachs2001use, rinaldi2013linguistic}.

Traditional language and motor rehabilitation programs heavily rely on auditory cues and are often implemented through repetitive, one-size-fits-all training protocols. Such methods lack interactivity and struggle to sustain children's motivation over extended periods. In contrast, emerging research indicates that multisensory feedback (integrating visual, auditory, and tactile cues) can significantly enhance cognitive engagement and promote neuroplasticity. For example, visual cues have been shown to improve postural control and motor timing \cite{cheng2004effects}, while vibrotactile feedback can directly convey speech rhythm or pitch cues through the skin, thereby compensating for auditory deficits \cite{kishon1996speechreading, cholewiak2003vibrotactile}.

With advancements in virtual reality (VR) technology and brain-computer interface (BCI) tools, opportunities to develop immersive, adaptive rehabilitation systems that provide real-time, multimodal feedback (personalized based on each child's performance) are increasing. However, the neural mechanisms underlying children's processing of these feedback combinations remain poorly understood, particularly in clinical pediatric populations. Functional near-infrared spectroscopy (fNIRS) offers a non-invasive method to observe cortical responses during natural tasks, providing valuable insights into attention, sensory integration, and motor planning in the developing brain \cite{lawrence2021evaluating, li2020effects, wang2021neural}.

This system aims to address this research gap by investigating the effects of different types and temporal sequences of multisensory feedback on behavioral outcomes and cortical activation patterns in children with hearing impairments. We developed a VR-based pitch-matching rehabilitation system that provides auditory, visual, and vibrotactile feedback, and utilized fNIRS to monitor activation in the prefrontal cortex, premotor area, and sensorimotor cortex regions. Through this system, we aim to uncover the relationship between real-time sensory feedback, user engagement, and neural plasticity. The findings of this study are expected to provide a basis for designing the next generation of personalized rehabilitation tools to better support language development, motor coordination, and autonomy in children with hearing impairments.

\section{\emph{MeloKids} WALKTHROUGH}
Our system is designed to support language and motor rehabilitation for children with hearing impairments by integrating real-time feedback from auditory, visual, and tactile channels. The system was developed using the Unity 3D engine and custom hardware, creating a training framework that maps pitch accuracy to multimodal outputs while measuring cortical responses via functional near-infrared spectroscopy (fNIRS). The training system is deployed on a virtual reality (VR) headset and synchronized with a vibrotactile feedback array and an audio processing unit. Its core comprises four main components: multisensory interaction, real-time feedback encoding, conditioned stimulus experimental control, and cognitive-neural response assessment.

\textbf{Multisensory Interaction.}
The core of the system is an interactive pitch matching task in which participants are guided to reproduce a series of musical fragments with predefined pitch curves through vocalization. These exercises are designed to simulate the melodic and rhythmic patterns associated with language learning. Microphones embedded in virtual reality devices capture voice input at a sampling rate of 10 kHz. The system uses an FFT analysis based on the cepstrum to estimate the fundamental frequency (F0) in real time.

Pitch data is then mapped to two types of sensory feedback. First, a visual overlay appears in the virtual reality environment, displaying a comparison between the real-time pitch trajectory and the target pitch curve. Second, the same pitch values trigger localized vibrations on a 1×7 cm wearable haptic display worn on the user's forearm or hand. The display consists of 18 piezoelectric actuators arranged in two columns of 32 rows, simulating a linear tactile pitch scale \cite{rogers2007choice, sakajiri2013accuracy}. By matching pitch with actuator position, children can perceive pitch height through physical tactile sensation, thereby forming an intuitive tactile association between vocalization and auditory perception \cite{kishon1996speechreading, lynch1992open}.

The integration of audio-visual-tactile modalities not only supports compensatory learning strategies but also encourages multi-channel engagement, potentially enhancing motivation, immersion, and sensorimotor skills in speech tasks \cite{woynaroski2013multisensory, shokur2018training}.

\textbf{Real-Time Feedback Encoding.}
To study the effects of feedback timing on user performance and cortical activation, the system supports two different feedback delivery modes: synchronous feedback and terminal feedback. In synchronous mode, participants receive real-time feedback immediately during vocalization: they hear the target melody, see a visual overlay of their pitch trajectory in the virtual reality environment, and feel vibrations corresponding to the wearable haptic array. This multimodal, synchronous feedback helps users correct themselves in real time, simulating the natural auditory-vocal feedback loop critical to the development of fluent language \cite{nesbitt2005framework, lalonde2020audiovisual}.

In delayed mode, feedback is delayed during vocalization and presented only after the vocalization segment is completed. Depending on specific conditions, users may receive auditory cues (confirmation sounds), visual assessments (scores or pitch trajectories), and haptic summaries (localized vibrations indicating accuracy of performance). This delayed structure encourages reflection on and internalization of vocal performance, enabling us to study retrospective awareness and cognitive load.

Each feedback channel (auditory, visual, and haptic) is precisely time-aligned and recorded, ensuring compatibility with neuroimaging data recorded via fNIRS. This synchronization enables us to map specific feedback events to cortical responses in motor and cognitive brain regions, facilitating a refined analysis of how time and modality interact to shape rehabilitation outcomes \cite{teo2016does, pomplun2022vibrotactile}.

\textbf{Self-Awareness Guidance.}
In parallel with neuroimaging studies, the system records detailed acoustic and behavioral metrics. Children's singing pitch is segmented, standardized, and compared with the target melody to calculate pitch deviation, pitch contour direction accuracy, and rhythmic precision \cite{mao2013acoustic}. After the session, participants complete the Game Engagement Questionnaire (GEQ) and the Intrinsic Motivation Inventory (IMI). A joint analysis of questionnaire scores, pitch data, and HbO values is conducted using analysis of variance (ANOVA) and Bonferroni correction to assess the impact of different sensory modalities on cognitive engagement and cortical plasticity during rehabilitation \cite{holper2009task, preacher2004spss}.

\section{FUTURE WORK}
Future work will focus on expanding the functionality and usability of the system. We plan to miniaturize the haptic feedback array into a more wearable form suitable for children and explore new feedback modalities, such as dynamic visual phoneme cues. The software will be expanded to support adaptive difficulty based on user performance and personalized settings. Ultimately, we aim to deploy the system in home and school environments to enable continuous, interactive rehabilitation beyond the laboratory.

\bibliographystyle{ACM-Reference-Format}
\bibliography{reference}

%%% -*-BibTeX-*-
%%% Do NOT edit. File created by BibTeX with style
%%% ACM-Reference-Format-Journals [18-Jan-2012].

\begin{thebibliography}{21}

%%% ====================================================================
%%% NOTE TO THE USER: you can override these defaults by providing
%%% customized versions of any of these macros before the \bibliography
%%% command.  Each of them MUST provide its own final punctuation,
%%% except for \shownote{} and \showURL{}.  The latter two
%%% do not use final punctuation, in order to avoid confusing it with
%%% the Web address.
%%%
%%% To suppress output of a particular field, define its macro to expand
%%% to an empty string, or better, \unskip, like this:
%%%
%%% \newcommand{\showURL}[1]{\unskip}   % LaTeX syntax
%%%
%%% \def \showURL #1{\unskip}           % plain TeX syntax
%%%
%%% ====================================================================

\ifx \showCODEN    \undefined \def \showCODEN     #1{\unskip}     \fi
\ifx \showISBNx    \undefined \def \showISBNx     #1{\unskip}     \fi
\ifx \showISBNxiii \undefined \def \showISBNxiii  #1{\unskip}     \fi
\ifx \showISSN     \undefined \def \showISSN      #1{\unskip}     \fi
\ifx \showLCCN     \undefined \def \showLCCN      #1{\unskip}     \fi
\ifx \shownote     \undefined \def \shownote      #1{#1}          \fi
\ifx \showarticletitle \undefined \def \showarticletitle #1{#1}   \fi
\ifx \showURL      \undefined \def \showURL       {\relax}        \fi
% The following commands are used for tagged output and should be
% invisible to TeX
\providecommand\bibfield[2]{#2}
\providecommand\bibinfo[2]{#2}
\providecommand\natexlab[1]{#1}
\providecommand\showeprint[2][]{arXiv:#2}

\bibitem[Amir et~al\mbox{.}(2003)]%
        {amir2003effect}
\bibfield{author}{\bibinfo{person}{Ofer Amir}, \bibinfo{person}{Noam Amir}, {and} \bibinfo{person}{Liat Kishon-Rabin}.} \bibinfo{year}{2003}\natexlab{}.
\newblock \showarticletitle{The effect of superior auditory skills on vocal accuracy}.
\newblock \bibinfo{journal}{\emph{The Journal of the Acoustical Society of America}} \bibinfo{volume}{113}, \bibinfo{number}{2} (\bibinfo{year}{2003}), \bibinfo{pages}{1102--1108}.
\newblock


\bibitem[Cheng et~al\mbox{.}(2004)]%
        {cheng2004effects}
\bibfield{author}{\bibinfo{person}{Pao-Tsai Cheng}, \bibinfo{person}{Chin-Man Wang}, \bibinfo{person}{Chia-Ying Chung}, {and} \bibinfo{person}{Chia-Ling Chen}.} \bibinfo{year}{2004}\natexlab{}.
\newblock \showarticletitle{Effects of visual feedback rhythmic weight-shift training on hemiplegic stroke patients}.
\newblock \bibinfo{journal}{\emph{Clinical rehabilitation}} \bibinfo{volume}{18}, \bibinfo{number}{7} (\bibinfo{year}{2004}), \bibinfo{pages}{747--753}.
\newblock


\bibitem[Cholewiak and Collins(2003)]%
        {cholewiak2003vibrotactile}
\bibfield{author}{\bibinfo{person}{Roger~W Cholewiak} {and} \bibinfo{person}{Amy~A Collins}.} \bibinfo{year}{2003}\natexlab{}.
\newblock \showarticletitle{Vibrotactile localization on the arm: Effects of place, space, and age}.
\newblock \bibinfo{journal}{\emph{Perception \& psychophysics}} \bibinfo{volume}{65}, \bibinfo{number}{7} (\bibinfo{year}{2003}), \bibinfo{pages}{1058--1077}.
\newblock


\bibitem[Holper et~al\mbox{.}(2009)]%
        {holper2009task}
\bibfield{author}{\bibinfo{person}{Lisa Holper}, \bibinfo{person}{Martin Biallas}, {and} \bibinfo{person}{Martin Wolf}.} \bibinfo{year}{2009}\natexlab{}.
\newblock \showarticletitle{Task complexity relates to activation of cortical motor areas during uni-and bimanual performance: a functional NIRS study}.
\newblock \bibinfo{journal}{\emph{Neuroimage}} \bibinfo{volume}{46}, \bibinfo{number}{4} (\bibinfo{year}{2009}), \bibinfo{pages}{1105--1113}.
\newblock


\bibitem[Kishon-Rabin et~al\mbox{.}(1996)]%
        {kishon1996speechreading}
\bibfield{author}{\bibinfo{person}{Liat Kishon-Rabin}, \bibinfo{person}{Arthur Boothroyd}, {and} \bibinfo{person}{Laurie Hanin}.} \bibinfo{year}{1996}\natexlab{}.
\newblock \showarticletitle{Speechreading enhancement: A comparison of spatial-tactile display of voice fundamental frequency (F 0) with auditory F 0}.
\newblock \bibinfo{journal}{\emph{The Journal of the Acoustical Society of America}} \bibinfo{volume}{100}, \bibinfo{number}{1} (\bibinfo{year}{1996}), \bibinfo{pages}{593--602}.
\newblock


\bibitem[Lachs et~al\mbox{.}(2001)]%
        {lachs2001use}
\bibfield{author}{\bibinfo{person}{Lorin Lachs}, \bibinfo{person}{David~B Pisoni}, {and} \bibinfo{person}{Karen~Iler Kirk}.} \bibinfo{year}{2001}\natexlab{}.
\newblock \showarticletitle{Use of audiovisual information in speech perception by prelingually deaf children with cochlear implants: A first report}.
\newblock \bibinfo{journal}{\emph{Ear and hearing}} \bibinfo{volume}{22}, \bibinfo{number}{3} (\bibinfo{year}{2001}), \bibinfo{pages}{236--251}.
\newblock


\bibitem[Lalonde and McCreery(2020)]%
        {lalonde2020audiovisual}
\bibfield{author}{\bibinfo{person}{Kaylah Lalonde} {and} \bibinfo{person}{Ryan~W McCreery}.} \bibinfo{year}{2020}\natexlab{}.
\newblock \showarticletitle{Audiovisual enhancement of speech perception in noise by school-age children who are hard of hearing}.
\newblock \bibinfo{journal}{\emph{Ear and Hearing}} \bibinfo{volume}{41}, \bibinfo{number}{4} (\bibinfo{year}{2020}), \bibinfo{pages}{705--719}.
\newblock


\bibitem[Lawrence et~al\mbox{.}(2021)]%
        {lawrence2021evaluating}
\bibfield{author}{\bibinfo{person}{Rachael~J Lawrence}, \bibinfo{person}{Ian~M Wiggins}, \bibinfo{person}{Jessica~C Hodgson}, {and} \bibinfo{person}{Douglas~EH Hartley}.} \bibinfo{year}{2021}\natexlab{}.
\newblock \showarticletitle{Evaluating cortical responses to speech in children: A functional near-infrared spectroscopy (fNIRS) study}.
\newblock \bibinfo{journal}{\emph{Hearing research}}  \bibinfo{volume}{401} (\bibinfo{year}{2021}), \bibinfo{pages}{108155}.
\newblock


\bibitem[Li et~al\mbox{.}(2020)]%
        {li2020effects}
\bibfield{author}{\bibinfo{person}{Qinbiao Li}, \bibinfo{person}{Jian Feng}, \bibinfo{person}{Jia Guo}, \bibinfo{person}{Zilin Wang}, \bibinfo{person}{Puhong Li}, \bibinfo{person}{Heshan Liu}, {and} \bibinfo{person}{Zhijun Fan}.} \bibinfo{year}{2020}\natexlab{}.
\newblock \showarticletitle{Effects of the multisensory rehabilitation product for home-based hand training after stroke on cortical activation by using NIRS methods}.
\newblock \bibinfo{journal}{\emph{Neuroscience letters}}  \bibinfo{volume}{717} (\bibinfo{year}{2020}), \bibinfo{pages}{134682}.
\newblock


\bibitem[Lynch et~al\mbox{.}(1992)]%
        {lynch1992open}
\bibfield{author}{\bibinfo{person}{Michael~P Lynch}, \bibinfo{person}{Rebecca~E Eilers}, {and} \bibinfo{person}{Patricia~J Pero}.} \bibinfo{year}{1992}\natexlab{}.
\newblock \showarticletitle{Open-set word identification by an adult with profound hearing impairment: Integration of touch, aided hearing, and speechreading}.
\newblock \bibinfo{journal}{\emph{Journal of Speech, Language, and Hearing Research}} \bibinfo{volume}{35}, \bibinfo{number}{2} (\bibinfo{year}{1992}), \bibinfo{pages}{443--449}.
\newblock


\bibitem[Mao et~al\mbox{.}(2013)]%
        {mao2013acoustic}
\bibfield{author}{\bibinfo{person}{Yitao Mao}, \bibinfo{person}{Mengchao Zhang}, \bibinfo{person}{Heather Nutter}, \bibinfo{person}{Yijing Zhang}, \bibinfo{person}{Qixin Zhou}, \bibinfo{person}{Qiaoyun Liu}, \bibinfo{person}{Weijing Wu}, \bibinfo{person}{Dinghua Xie}, {and} \bibinfo{person}{Li Xu}.} \bibinfo{year}{2013}\natexlab{}.
\newblock \showarticletitle{Acoustic properties of vocal singing in prelingually-deafened children with cochlear implants or hearing aids}.
\newblock \bibinfo{journal}{\emph{International Journal of Pediatric Otorhinolaryngology}} \bibinfo{volume}{77}, \bibinfo{number}{11} (\bibinfo{year}{2013}), \bibinfo{pages}{1833--1840}.
\newblock


\bibitem[Nesbitt(2005)]%
        {nesbitt2005framework}
\bibfield{author}{\bibinfo{person}{Keith Nesbitt}.} \bibinfo{year}{2005}\natexlab{}.
\newblock \showarticletitle{A framework to support the designers of haptic, visual and auditory displays}. In \bibinfo{booktitle}{\emph{Guidance on Tactile and Haptic Interactions}}. University of Saskatoon, \bibinfo{pages}{54--64}.
\newblock


\bibitem[Pomplun et~al\mbox{.}(2022)]%
        {pomplun2022vibrotactile}
\bibfield{author}{\bibinfo{person}{Ella Pomplun}, \bibinfo{person}{Ashiya Thomas}, \bibinfo{person}{Erin Corrigan}, \bibinfo{person}{Valay~A Shah}, \bibinfo{person}{Leigh~A Mrotek}, {and} \bibinfo{person}{Robert~A Scheidt}.} \bibinfo{year}{2022}\natexlab{}.
\newblock \showarticletitle{Vibrotactile perception for sensorimotor augmentation: Perceptual discrimination of vibrotactile stimuli induced by low-cost eccentric rotating mass motors at different body locations in young, middle-aged, and older adults}.
\newblock \bibinfo{journal}{\emph{Frontiers in Rehabilitation Sciences}}  \bibinfo{volume}{3} (\bibinfo{year}{2022}), \bibinfo{pages}{895036}.
\newblock


\bibitem[Preacher and Hayes(2004)]%
        {preacher2004spss}
\bibfield{author}{\bibinfo{person}{Kristopher~J Preacher} {and} \bibinfo{person}{Andrew~F Hayes}.} \bibinfo{year}{2004}\natexlab{}.
\newblock \showarticletitle{SPSS and SAS procedures for estimating indirect effects in simple mediation models}.
\newblock \bibinfo{journal}{\emph{Behavior research methods, instruments, \& computers}} \bibinfo{volume}{36}, \bibinfo{number}{4} (\bibinfo{year}{2004}), \bibinfo{pages}{717--731}.
\newblock


\bibitem[Rinaldi et~al\mbox{.}(2013)]%
        {rinaldi2013linguistic}
\bibfield{author}{\bibinfo{person}{Pasquale Rinaldi}, \bibinfo{person}{Francesca Baruffaldi}, \bibinfo{person}{Sandro Burdo}, {and} \bibinfo{person}{Maria~Cristina Caselli}.} \bibinfo{year}{2013}\natexlab{}.
\newblock \showarticletitle{Linguistic and pragmatic skills in toddlers with cochlear implant}.
\newblock \bibinfo{journal}{\emph{International Journal of Language \& Communication Disorders}} \bibinfo{volume}{48}, \bibinfo{number}{6} (\bibinfo{year}{2013}), \bibinfo{pages}{715--725}.
\newblock


\bibitem[Rogers(2007)]%
        {rogers2007choice}
\bibfield{author}{\bibinfo{person}{Charles~H Rogers}.} \bibinfo{year}{2007}\natexlab{}.
\newblock \showarticletitle{Choice of stimulator frequency for tactile arrays}.
\newblock \bibinfo{journal}{\emph{IEEE Transactions on man-machine systems}} \bibinfo{volume}{11}, \bibinfo{number}{1} (\bibinfo{year}{2007}), \bibinfo{pages}{5--11}.
\newblock


\bibitem[Sakajiri et~al\mbox{.}(2013)]%
        {sakajiri2013accuracy}
\bibfield{author}{\bibinfo{person}{Masatsugu Sakajiri}, \bibinfo{person}{Shigeki Miyoshi}, \bibinfo{person}{Kenryu Nakamura}, \bibinfo{person}{Satoshi Fukushima}, {and} \bibinfo{person}{Tohru Ifukube}.} \bibinfo{year}{2013}\natexlab{}.
\newblock \showarticletitle{Accuracy of voice pitch control in singing using tactile voice pitch feedback display}. In \bibinfo{booktitle}{\emph{2013 IEEE International Conference on Systems, Man, and Cybernetics}}. IEEE, \bibinfo{pages}{4201--4206}.
\newblock


\bibitem[Shokur et~al\mbox{.}(2018)]%
        {shokur2018training}
\bibfield{author}{\bibinfo{person}{Solaiman Shokur}, \bibinfo{person}{Ana~RC Donati}, \bibinfo{person}{Debora~SF Campos}, \bibinfo{person}{Claudia Gitti}, \bibinfo{person}{Guillaume Bao}, \bibinfo{person}{Dora Fischer}, \bibinfo{person}{Sabrina Almeida}, \bibinfo{person}{Vania~AS Braga}, \bibinfo{person}{Patricia Augusto}, \bibinfo{person}{Chris Petty}, {et~al\mbox{.}}} \bibinfo{year}{2018}\natexlab{}.
\newblock \showarticletitle{Training with brain-machine interfaces, visuo-tactile feedback and assisted locomotion improves sensorimotor, visceral, and psychological signs in chronic paraplegic patients}.
\newblock \bibinfo{journal}{\emph{PloS one}} \bibinfo{volume}{13}, \bibinfo{number}{11} (\bibinfo{year}{2018}), \bibinfo{pages}{e0206464}.
\newblock


\bibitem[Teo et~al\mbox{.}(2016)]%
        {teo2016does}
\bibfield{author}{\bibinfo{person}{Wei-Peng Teo}, \bibinfo{person}{Makii Muthalib}, \bibinfo{person}{Sami Yamin}, \bibinfo{person}{Ashlee~M Hendy}, \bibinfo{person}{Kelly Bramstedt}, \bibinfo{person}{Eleftheria Kotsopoulos}, \bibinfo{person}{Stephane Perrey}, {and} \bibinfo{person}{Hasan Ayaz}.} \bibinfo{year}{2016}\natexlab{}.
\newblock \showarticletitle{Does a combination of virtual reality, neuromodulation and neuroimaging provide a comprehensive platform for neurorehabilitation?--a narrative review of the literature}.
\newblock \bibinfo{journal}{\emph{Frontiers in human neuroscience}}  \bibinfo{volume}{10} (\bibinfo{year}{2016}), \bibinfo{pages}{284}.
\newblock


\bibitem[Wang et~al\mbox{.}(2021)]%
        {wang2021neural}
\bibfield{author}{\bibinfo{person}{Yuyang Wang}, \bibinfo{person}{Lili Liu}, \bibinfo{person}{Ying Zhang}, \bibinfo{person}{Chaogang Wei}, \bibinfo{person}{Tianyu Xin}, \bibinfo{person}{Qiang He}, \bibinfo{person}{Xinlin Hou}, {and} \bibinfo{person}{Yuhe Liu}.} \bibinfo{year}{2021}\natexlab{}.
\newblock \showarticletitle{The neural processing of vocal emotion after hearing reconstruction in prelingual deaf children: a functional near-infrared spectroscopy brain imaging study}.
\newblock \bibinfo{journal}{\emph{Frontiers in neuroscience}}  \bibinfo{volume}{15} (\bibinfo{year}{2021}), \bibinfo{pages}{705741}.
\newblock


\bibitem[Woynaroski et~al\mbox{.}(2013)]%
        {woynaroski2013multisensory}
\bibfield{author}{\bibinfo{person}{Tiffany~G Woynaroski}, \bibinfo{person}{Leslie~D Kwakye}, \bibinfo{person}{Jennifer~H Foss-Feig}, \bibinfo{person}{Ryan~A Stevenson}, \bibinfo{person}{Wendy~L Stone}, {and} \bibinfo{person}{Mark~T Wallace}.} \bibinfo{year}{2013}\natexlab{}.
\newblock \showarticletitle{Multisensory speech perception in children with autism spectrum disorders}.
\newblock \bibinfo{journal}{\emph{Journal of autism and developmental disorders}} \bibinfo{volume}{43}, \bibinfo{number}{12} (\bibinfo{year}{2013}), \bibinfo{pages}{2891--2902}.
\newblock


\end{thebibliography}

\end{document}